\documentclass[aip,cha,amsmath,amssymb,preprint]{revtex4-1}

\usepackage{mathptmx}
\usepackage{color}
\usepackage{pifont, amsthm}
\usepackage{graphicx}

\begin{document}
	
	\title{Stability of entrainment of a continuum of coupled oscillators}
	
	\author{Jordan Snyder}
	\email{jasnyder@math.ucdavis.edu}\affiliation{Department of Mathematics, University of California, Davis, CA 95616}

	\author{Anatoly Zlotnik}
	\affiliation{Applied Mathematics and Plasma Physics (T-5), Theoretical Division, Los Alamos National Laboratory, Los Alamos, NM 87544}
	
	\author{Aric Hagberg}
	\affiliation{Applied Mathematics and Plasma Physics (T-5), Theoretical Division, Los Alamos National Laboratory, Los Alamos, NM 87544}
	
	\begin{abstract}
		Complex natural and engineered systems are ubiquitous and their behavior is challenging to characterize and control. We examine the design of the entrainment process for an uncountably infinite collection of coupled phase oscillators that are all subject to the same periodic driving signal.  In the absence of coupling, an appropriately designed input can result in each oscillator attaining the frequency of the driving signal, with a phase offset determined by its natural frequency.  We consider a special case of interacting oscillators in which the coupling tends to destabilize the phase configuration to which the driving signal would send the collection in the absence of coupling. In this setting we derive stability results that characterize the trade-off between the effects of driving and coupling, and compare these results to the well-known Kuramoto model of a collection of free-running coupled oscillators.
	\end{abstract}
	
	
	\maketitle
	
	\begin{quotation}
		Two well-understood approaches can be applied to impose coherent behavior in a diverse population of dynamical systems: the ``top-down'' approach of applying a common driving signal, and the ``bottom-up'' approach of imposing pairwise coupling. While these  approaches yield similar behaviors, their precise characteristics can put them in opposition. In this article we study a situation that highlights both the synergy and tension that can exist between driving and coupling in collections of oscillators.
	\end{quotation}
	
	\section{Introduction}
	
	A growing variety of collective dynamic behaviors in networked periodic phenomena are observed across disciplines, with many well-known examples such as circadian cycles, injection-locked semiconductors, and emerging applications including battery charging cycles and neural information processing.  Analysis, design, and control of interacting collections of rhythmic or oscillatory processes whose scale or complexity is beyond the scope of classical dynamical systems theory requires new mathematical frameworks.  Developing continuum approximations to oscillatory phenomena over large-scale networks will create paths towards practical solutions to motivating applications.
	
	One basic class of phenomena that is of interest in chronobiology \cite{prolo2005circadian}, electrochemistry \cite{kiss02,kiss07}, neuroscience \cite{uhlhaas06,rodriguez1999perception}, and  power grid engineering \cite{dorfler13} is the formation of coherent behavior in collections of interacting units.  Such coherence can be imposed from outside, or can arise through the intrinsic interactions themselves.  There is a long history of studying the emergence of coherent motion in oscillators using phase model representations, dating back to Winfree \cite{winfree1967biological,winfree2001geometry} and Kuramoto \cite{kuramoto1975self,kuramoto2012chemical}.  The canonical examples that were posed in these early studies have been widely examined in subsequent work \cite{brown2003globally}, because they exhibit a rich phenomenology while admitting beautiful mathematical descriptions within an extensive range of analytical settings. While original studies focused on mutual entrainment \cite{kuramoto1975self}, in which coherent motion arises purely from interactions between individual units, more recent studies have investigated the effect of externally-imposed coherence in the form of external driving.
	
	A pioneering study of forced, coupled oscillations was performed by Sakaguchi \cite{sakaguchi1988}, who considered an infinite, heterogeneous, population of oscillators subject to global sinusoidal coupling and uniform sinusoidal forcing. By deriving a self-consistent equation for the order parameter measuring phase alignment, Sakaguchi was able to predict transitions between regimes of incoherence, mutual entrainment, and forced entrainment. These transitions were subsequently investigated in more detail by Antonsen and collaborators using a linear stability approach \cite{antonsen2008external}. The detailed nature of the bifurcations remained elusive, and suggested an underlying two-dimensional structure which had yet to be exploited.  This two-dimensional structure was indeed discovered by Ott and Antonsen in their seminal work \cite{ott2008low}, which uncovered a particular low-dimensional manifold that captures much of the asymptotic behavior of a wide family of models of coupled phase oscillators; in particular, the forced Kuramoto model was identified as a possible application of this dimension reduction. Subsequent work has shown that, under certain mild conditions, this manifold is globally attractive \cite{ott2009, ott2011}. This reduction represents an enormous simplification, because in many cases it permits closed-form evolution equations for the synchrony order parameter directly.  Building on the framework established by Ott and Antonsen \cite{ott2008low}, Childs and Strogatz \cite{childs2008stability} were able study the dynamics of the forced Kuramoto model on the two-dimensional attractive manifold, and found a complete picture of a system's bifurcation structure. It should be noted that studies of the effect of forcing in this context have almost exclusively considered a sinusoidal forcing function, due to the analytical tractability of the resulting model. Recently, complementary work has been done that examines the role of random forcing applied to a population of sinusoidally-coupled oscillators\cite{pimenova2016interplay}.
	
	Beyond characterizing the phenomenology of natural and engineered complex oscillating systems, emerging applications in neural systems, electrochemistry, and power grid engineering require new capabilities to control and manipulate the behavior of such phenomena.  Indeed, the ability to control a system is the ultimate validation of our understanding of its behavior.  For oscillating systems, a general picture of frequency modulation by external forcing was first laid out in 1946 by Adler \cite{adler1946}, who derived equations describing the amount by which an external drive signal can shift an oscillator's frequency and amplitude. The idea of ``injection locking'' has since been of major importance in many fields of engineering \cite{razavi2003, barnes2011}.  One prediction made by Adler was that an oscillator driven at a frequency different from its own may lock to the driving frequency, and exhibit a phase shift relative to the drive signal which is determined by its natural frequency. For the simple case considered originally, this function is sinusoidal.  However, for a general forcing function and a general phase oscillator higher harmonics may be present, as seen in experiments and derived analytically \cite{hunter03,zlotnik2014optimal}. The general framework of using periodic forcing signals to control the entrainment of nonlinear oscillating systems has been exploited to explore energy- and time-optimal control strategies for entrainment of one or more phase oscillators  \cite{efimov09,zlotnik2012,zlotnik2013}. The effect of coupling on the efficacy of these control strategies remains unexplored.
	
	A challenge in specifying the forcing input to control collections of coupled oscillators is that they are underactuated; the entire collection of similar dynamical systems with possibly complicated individual behavior must be controlled using a small number of inputs.  To overcome this challenge we observe that the entrained or coherent state of a controlled collection of oscillators is characterized not only by synchronization to a forcing frequency, but also by the distribution of subsystems on the neighborhood of a nominal periodic orbit.  For a finite collection, it is possible to construct a forcing signal to achieve precise control of the relative phases of an ensemble of structurally similar oscillators with slight heterogeneity in frequencies \cite{zlotnik2016phase}.  With the understanding that such ``phase-selective control'' is possible for small, finite collections, we examine how the mathematical framework can be extended to continuum systems.  Further, we examine the effect of coupling between subsystems, which tends to drive phase differences to zero.
	
	In this paper we explore a continuum approximation of a very large collection of coupled oscillators subject to a common periodic (but non-sinusoidal) forcing, so that both coupling and forcing influence the collective behavior. Specifically, we consider a situation in which the forcing drives individual phases to be maximally different (in a certain precise sense), while the coupling tends to align the phases. To quantify the trade-off between these two effects, we compute, as a function of the coupling strength, the asymptotic stability of a fixed point in which the phases show no global alignment. By finding the critical coupling strength above which this fixed point is unstable, we demonstrate that mutual synchronization of entrained coupled oscillators occurs \emph{before} mutual synchronization of unforced coupled oscillators, despite the imposed diversity of phases. Moreover, numerical experiments confirm that the external forcing has facilitated phase alignment which is \emph{greater} than that in the unforced case. Our results demonstrate that measuring only phase alignment is bound to miss important information about the global organization of a population of oscillators.

	\section{Preliminaries}
	
	\subsection{Entrainment of oscillators}
	
	We next describe how a heterogeneous population of oscillators can be caused to move at a single frequency by application of a suitable forcing function. Mathematical details can be found in several standard references\cite{strogatz2001nonlinear,hoppensteadt2012weakly,winfree1967biological}.
	
	As our basic model of an oscillator, we take a \emph{phase model}, first popularized by Winfree\cite{winfree1967biological}. For $i=1,\dots, N$, the $i^{\text{th}}$ oscillator is described by the ODE
	\begin{equation}
	\dot{\psi}_i = \omega_i + Z(\psi_i)u \label{eq:PRC_dynamics}\,,
	\end{equation}
	where $\psi_i \in [0,2\pi)$ is the phase, `` $\dot{}$ '' denotes the derivative with respect to time, $\omega_i\in\mathbb{R}$ is the natural frequency, $u=u(t)$ is an external forcing, and $Z(\psi_i)$ is known as the \emph{phase response curve}, or PRC. The PRC determines the change in phase resulting from an infinitesimal external force applied at a given phase on the limit cycle \cite{nakao2016phase,ermentrout1996type}. The equation (\ref{eq:PRC_dynamics}) can be derived by considering the lowest-order approximation of the effect of an external force acting on a system near a stable limit cycle\cite{kuramoto1975self,schwemmer2012theory}, and in this sense is representative of a wide class of forced periodic motions.
	
	A standard approach to analyzing entrainment is to take $u(t) = v(\Omega t)$, where $v$ has period $2\pi$ so that $\Omega$ denotes the (angular) frequency of the driving signal. If $\Omega$ is not too far from $\omega_i$, we suppose that $\psi_i$ will behave as $\Omega t$, plus a slowly-varying phase offset. We formalize this supposition by making the change of coordinates $\psi_i = \Omega t + \phi_i$, where $\phi_i$ now represents the phase offset. In the $\phi_i$ coordinate system, the dynamics now read
	\begin{equation}
	\dot{\phi}_i = \Delta\omega_i  + Z(\phi_i + \Omega t)v(\Omega t), \label{eq:PRC_dynamics_shifted}
	\end{equation}
	where we have introduced the \emph{frequency detuning} $\Delta\omega_i \equiv \omega_i - \Omega$. Finally it is possible to approximate (\ref{eq:PRC_dynamics_shifted}) by the time-averaged system~\cite{hoppensteadt2012weakly}
	\begin{equation}
	\dot{\varphi}_i = \Delta\omega_i + \Lambda_v(\varphi_i), \label{eq:PRC_dynamics_averaged}
	\end{equation}
	where we have introduced the \emph{interaction function}
	\begin{equation}
	\Lambda_v(\varphi) = \frac{1}{2\pi} \intop_{0}^{2\pi}Z(\varphi+\theta)v(\theta)d\theta,
	\end{equation}
	in the sense that there exists a change of variables $\varphi_i = \phi_i + h(\varphi_i,\phi_i)$ that maps solutions of (\ref{eq:PRC_dynamics_shifted}) to those of (\ref{eq:PRC_dynamics_averaged}).
	
	If the frequency detunings $\{\Delta\omega_i\}$ are such that (\ref{eq:PRC_dynamics_averaged}) has a stable fixed point solution for all $i=1,\dots,N$, then the phases of all oscillators will be constant in the moving reference frame with frequency $\Omega$. In other words, the entire population can be \emph{entrained} by the driving signal $u=v(\Omega t)$.
	
	Despite having equal frequencies, the oscillators will, in general, have different phases, since the solutions to the fixed point equation $\Delta\omega_i + \Lambda_v(\varphi_i) = 0$ depend on the value of $\Delta\omega_i$. This fact can be exploited to design a forcing function that elicits frequency locking with a known distribution of phases, irrespective of initial conditions\cite{zlotnik2016phase}.

	\subsection{The Kuramoto Model}
	\label{sec:kuramoto_intro}
	To frame our study of phase coupling, we discuss some standard methods and results relating to synchronization of phase oscillators. Kuramoto introduced a model of the form
	\begin{equation}
	\dot{\varphi_i} = \omega_i + \frac{K}{N}\sum_{j=1}^{N} \sin(\varphi_j - \varphi_i)\,, \label{eq:kuramoto_fd}
	\end{equation}
	which was derived as a ``simplest'' model for a collection of self-sustained linearly-coupled oscillators \cite{kuramoto1975self}. Here $\{ \varphi_i\}$ are the phases of $N$ oscillators, $\{\omega_i\}$ their natural frequencies (which we allow to take any real values), and $K>0$ is the strength of coupling.
	
	This ODE can be instructively rewritten in the form
	\begin{equation}
	\dot{\varphi_i} = \omega_i + KR\sin(\Phi - \varphi_i)\,,
	\end{equation}
	where we have used the \emph{synchrony} $R\in[0,1]$, and the \emph{average phase} $\Phi\in[0,2\pi)$, first introduced by Kuramoto \cite{kuramoto1975self} and defined by the formula
	\begin{equation}
	Re^{i\Phi} = \frac{1}{N}\sum_{j=1}^{N} e^{i\varphi_j}. \label{eq:order_param}
	\end{equation}
	In this sense, this form of coupling is \emph{mean-field} in character, as each phase feels a force determined by an average over the entire population.
	
	The key features of this model are
	\begin{enumerate}
		\item The oscillators have differing intrinsic frequency: $\omega_i \ne \omega_j$ \,,
		\item The coupling tends to drive phases towards the mean (provided $K>0$, which we assume throughout).
	\end{enumerate}
	
	These two features are at odds with each other, and they undergo a trade-off at a critical value of the coupling strength, $K=K_c^{\text{unf}}$ (where we use the superscript "unf" to emphasize that this is the critical coupling strength in the \emph{unforced} case). If $K<K_c^{\text{unf}}$ the population of oscillators does not show global alignment towards any particular phase, while for $K>K_c^{\text{unf}}$, this situation breaks down and a subset of the oscillators attains the same frequency and group together in phase, establishing a preferred direction and a nonzero value of the synchrony $R$.
	
	To make these statements precise it is useful to consider a \emph{mean-field approximation}. We suppose that the population of oscillators is large enough that averaging over this population is well approximated by averaging over a probability distribution that describes the behavior of a typical oscillator. General background on the mean field Kuramoto model can be found in various review articles \cite{strogatz2000kuramoto,Acebron2005}.
	
	The main result we quote from the extensive body of literature on the Kuramoto model is that in the limit of $N\to \infty$, if the oscillators' natural frequencies are drawn at random from a probability distribution having density $g(\omega)$, unimodal and symmetric about zero, then the critical coupling strength described above is given by
	\begin{equation}
	\label{eq:unforced_kc}
	K_c^{\text{unf}} = \frac{2}{\pi g(0)}.
	\end{equation}
	
	The expression (\ref{eq:unforced_kc}) can be taken as a precise quantification of the trade-off between intrinsic disorder ($g(0)$) and coupling ($K$). The possibly surprising fact that $K_c^{\text{unf}}$ depends \emph{only} on the value of $g$ at the center of the distribution, and no other features of this distribution, is because the first oscillators to synchronize are those whose natural frequencies lie at the center of the distribution. The rest of the density $g$ then determines the growth of $R$ with $K>K_c^{\text{unf}}$.
	
	In what follows, we will define a new model, show that it exhibits behavior that is qualitatively similar to that of the Kuramoto model, and find the location of the corresponding critical point. The expression (\ref{eq:unforced_kc}) will serve as reference to interpret our results.

	\section{Model for Forcing of Coupled Oscillations}
	
	\subsection{Finite $N$}
	
	We now formulate a model of a population of oscillators that exhibits both frequency alignment by broadcast forcing and phase alignment by attractive coupling. Many similar models have been developed \cite{sakaguchi1988,mirollo1990a,childs2008stability,ott2008low,antonsen2008external}, and our present formulation aims to augment the rich existing literature.
	
	In general, we can consider each oscillator to respond to external forcing according to one phase response curve, and to respond to forcing from its neighboring oscillators according to another phase response curve. That is,
	\begin{equation}
	\dot{\psi_i} = \omega_i + Z_e(\psi_i)u(t) + \frac{K}{N}\sum_{j=1}^{N}Z_c(\psi_i) f(\psi_j)
	\end{equation}
	where $Z_e$ is the PRC for external forcing, $Z_c$ is the PRC for coupling, and $f()$ describes the force an oscillator exerts on its neighbors as a function of its phase. The prefactor $K/N$ allows us to adjust the coupling strength $K$ in a way that allows comparison between different values of $N$.
	
	Assuming, as before, that $u(t) = v(\Omega t)$ with $v$ having period $2\pi$, we move into a rotating reference frame with frequency $\Omega$ and average over one period of the driving signal, obtaining the averaged equations
	\begin{equation}
	\dot{\varphi_i} = \Delta \omega_i + \Lambda_v(\varphi_i) + \frac{K}{N}\sum_{j=1}^{N} g(\varphi_j -\varphi_i)
	\end{equation}
	where $\varphi_i, \Delta\omega_i$, and $\Lambda_v$ are defined as before (\ref{eq:PRC_dynamics_averaged}) and $g(\Delta\varphi) =(2\pi)^{-1} \intop_{0}^{2\pi} Z_c(\theta + \Delta\varphi) f(\theta) d\theta$.
	
	Clearly, many different systems may be defined in this form given appropriate choices for $Z_e$, $Z_c$, $v$, and $f$. In order to exhibit the qualitative features of phase dispersion caused by external forcing combined with phase alignment caused by coupling, while retaining tractability, we assume that $Z_c$ and $f$ are such that $g(\Delta\varphi)=\sin(\Delta\varphi)$.
	
	Hence, we take a model of the form
	\begin{equation}
	\dot{\varphi_i} = \Delta\omega_i + \Lambda_v(\varphi_i) + \frac{K}{N}\sum_{j=1}^{N} \sin(\varphi_j-\varphi_i), \label{eq:dynamics}
	\end{equation}
	which can also be written
	\begin{equation}
	\dot{\varphi_i} = \Delta\omega_i + \Lambda_v(\varphi_i) + KR\sin(\Phi-\varphi_i),
	\end{equation}
	with $R$ and $\Phi$ defined as in (\ref{eq:order_param}).
	
	As a first step, we choose $\{\Delta\omega_i\}$ and $\Lambda_v$ such that all oscillators can be entrained individually, but the resulting phase offsets are as far as possible from alignment. This can be achieved by setting
	\begin{equation}
	\Delta\omega_i = \frac{2i}{N}-1 \,,
	\label{eq:detunings}
	\end{equation}
	and
	\begin{equation} \label{eq:sawtooth_interaction_function}
	\Lambda_v(\varphi)=\frac{-\varphi}{\pi},\qquad \varphi\in (-\pi, \pi].
	\end{equation}
	We refer to the function defined in (\ref{eq:sawtooth_interaction_function}) as the \emph{sawtooth interaction function}, or just \emph{sawtooth}, as it has a sawtooth shape when plotted on $\mathbb{R}$.
	
	The standard unforced Kuramoto model with this choice of natural frequencies has been recently studied by Ottino-L\"{o}ffler and Strogatz \cite{ottino-loffler2016}, who found the asymptotic behavior of the locking threshold as $N\to\infty$, in agreement with results in the thermodynamic limit obtained earlier by Paz\'{o} \cite{pazo2005}. These results will serve as a reference to put our findings in context. For now, we return to the forced case.
	
	In the absence of coupling ($K=0$), the $i^{\text{th}}$ oscillator will be driven to a phase offset $\varphi_i^*$ defined by
	\begin{equation}
	\label{eq:finite_N_fixed_point}
	\Delta\omega_i + \Lambda_v(\varphi_i^*) = 0 \implies \varphi_i^* = \pi\Delta\omega_i = \frac{2\pi i}{N} - \pi.
	\end{equation}
	A straightforward calculation shows that for this phase configuration, the synchrony is $R=0$. For this reason, we refer to this fixed point as the \emph{desynchronized state}. Another term used to describe such a state is \emph{splay state}. The point $\varphi^*=(\varphi_i^*)\in (-\pi,\pi]^N$ is a fixed point of the dynamics~(\ref{eq:dynamics}) for any value of coupling strength $K$.
	
	In this respect, the situation is similar to the \emph{incoherent state} discussed for the Kuramoto model in Section~\ref{sec:kuramoto_intro}, with the key difference that in this case, all oscillators have attained identical frequency locking to the forcing input. We proceed to study the asymptotic stability of this fixed point as a function of $K$, and obtain a critical coupling strength $K_c$ analogous to $K_c^{\text{unf}}$ as defined in ~(\ref{eq:unforced_kc}).
	
	\subsection{The $N\to\infty$ limit}
	
	Next we introduce a thermodynamic limit of the model~(\ref{eq:dynamics}), and the fixed point corresponding to that defined in~(\ref{eq:finite_N_fixed_point}).
	
	We replace our population of oscillators, formerly a collection of $N$ individual oscillators with natural frequencies evenly spaced from $-1$ to $1$,  by a continuum of oscillators with natural frequencies distributed uniformly on $[-1,1]$.
	
	Because our state of interest for finite $N$ is such that each oscillator's phase is fixed at a value determined by its natural frequency, we describe the state of our infinite system by a function $\varphi(\omega)$ that gives the phase of any oscillator having natural frequency $\omega$. As the system evolves the whole function $\varphi(\omega)$ will change in time, but for visual clarity we omit writing the time-dependence explicitly when discussing fixed points. This sort of formulation is used, for example, by Mirollo and Strogatz \cite{mirollo1990a}, except that oscillators are indexed by their ``pinning phase'' rather than their natural frequency. We describe this work in more detail in Section \ref{sec:relation_to_previous_work}.
	
	To determine fixed points, we must establish the dynamics in the appropriate continuum setting.  The intrinsic dynamics and effects of forcing remain the same, so we only need to concern ourselves with the coupling term. For finite $N$, we simply had an average over the population, and in the infinite setting, we use a mean-field approach to say that averaging over the infinite population is equivalent to averaging over the distribution of natural frequencies\cite{strogatz2000kuramoto}. Our infinite-dimensional dynamics are
	\begin{equation}
	\label{eq:infinite_dim_dynamics}
	\partial_t \varphi(\omega) = \omega + \Lambda_v(\varphi(\omega)) + K\intop_{\mathbb{R}} g(\omega') \sin(\varphi(\omega')-\varphi(\omega)) d\omega' \,,
	\end{equation}
	where $g$ is the density of the distribution of natural frequencies. These dynamics can be rewritten in the form
	\begin{equation}
	\label{eq:infinite_dim_dynamics_mf}
	\partial_t \varphi(\omega) = \omega + \Lambda_v(\varphi(\omega)) + KR\sin(\Phi - \varphi(\omega)) \,,
	\end{equation}
	where $R$ and $\Phi$ are the synchrony and average phase, defined for the infinite system as
	\begin{equation}
	Re^{i\Phi} = \intop_{\mathbb{R}} g(\omega)e^{i\varphi(\omega)} d\omega.
	\end{equation}
	
	Using the sawtooth interaction function introduced above (see (\ref{eq:sawtooth_interaction_function})), and $g(\omega) = 1/2$ for $\omega\in[-1,1]$ and $0$ elsewhere, the fixed point condition for $\varphi$ now reads
	\begin{equation}
	\label{eq:infinite_dim_fixed_point_condition}
	0 = \omega - \frac{\varphi(\omega)}{\pi} +K\intop_{-1}^{1} \frac{1}{2}\sin(\varphi(\omega')-\varphi(\omega))d\omega' \,,
	\end{equation}
	
	A straightforward calculation shows that the function $\varphi(\omega)=\pi\omega$ satisfies the condition (\ref{eq:infinite_dim_fixed_point_condition}). Note that this is precisely the infinite-$N$ analog of the finite-$N$ fixed point defined in (\ref{eq:finite_N_fixed_point}). In what follows, we perform a linear stability analysis, finding the coupling strength $K_c$ at which this state becomes unstable.

	\section{Stability Analysis of the Entrainment Phase Distribution}
	\subsection{Finite $N$}
	We now analyze the stability of the fixed point $\varphi^*=(\varphi_i^*)\in (-\pi,\pi]^N$ as defined in equation (\ref{eq:finite_N_fixed_point}).
	
	Asymptotic stability of $\varphi^*$ is controlled by the spectrum $\sigma({\bf J})$ of the Jacobian ${\bf J}$ of the right-hand side of (\ref{eq:dynamics}) with respect to $\varphi$, evaluated at $\varphi^*$. If every element of $\sigma({\bf J})$ has negative real part, then $\varphi^*$ is an asymptotically stable fixed point, and if any element of $\sigma({\bf J})$ has positive real part, then $\varphi^*$ is unstable\cite{strogatz2001nonlinear}. The matrix elements of ${\bf J}$ are
	\begin{equation}
	J_{ij}=\left(\Lambda'_{v}(\varphi_{i}^{*})-\frac{K}{N}\sum_{k\ne i}\cos(\varphi_{i}^{*}-\varphi_{k}^{*})\right)\delta_{ij} +(1-\delta_{ij})\frac{K}{N}\cos\left(\varphi_{i}^{*}-\varphi_{j}^{*}\right) \,,
	\end{equation}
	where $\delta_{ij}$ is the Kronecker delta. A straightforward calculation (see Appendix~\ref{sec:fin_dim_jac}) shows that
	\begin{equation}
	\label{eq:spectrum_of_jac_fd}
	\sigma(\mathbf{J})=\left\{ \frac{-1}{\pi},\frac{-1}{\pi}+\frac{K}{2}\right\}\,,
	\end{equation}
	so the desynchronized state has a critical point $K_c = 2/\pi$ and
	is linearly stable when $K<2/\pi$, and linearly unstable for $K>2/\pi$.
	
	\subsection{The $N\to\infty$ limit}
	
	Finally we will perform a linear stability analysis of the desynchronized fixed point of the infinite-$N$ model (\ref{eq:infinite_dim_dynamics}). For details of the calculation presented below, see Appendix~\ref{sec:infinite_dim_linearize}.
	
	To obtain a linearization of the dynamics near the fixed point $\varphi^*(\omega)=\pi\omega$, we consider an infinitesimal perturbation,
	\begin{equation}
	\varphi(\omega) = \varphi^*(\omega) + \epsilon\eta(\omega) \,,
	\end{equation}
	where $0<\epsilon\ll 1$ and $\eta\colon [-1,1]\to\mathbb{R}$ is a function which we take to be bounded and measurable.
	
	Inserting this form into (\ref{eq:infinite_dim_dynamics}) and collecting terms by order of $\epsilon$ yields
	\begin{align}
	\mathcal{O}(\epsilon^0) &\colon \quad \partial_t\varphi^*(\omega) = \omega - \frac{\varphi^*(\omega)}{\pi} \nonumber + K\intop_{-1}^{1}\frac{1}{2} \sin(\varphi^*(\omega')-\varphi^*(\omega)) d\omega' \nonumber \\
	\mathcal{O}(\epsilon^1) &\colon \quad \partial_t\eta(\omega) = -\frac{1}{\pi}\eta(\omega) + K\intop_{-1}^{1} \frac{1}{2} \cos(\varphi^*(\omega')-\varphi^*(\omega))\eta(\omega') d\omega'. \label{eq:infinite_dim_linearized_dynamics}
	\end{align}
	As expected, the $\mathcal{O}(\epsilon^0)$ equation holds by the fact that $\varphi^*$ is a fixed point, and the $\mathcal{O}(\epsilon^1)$ gives the time evolution of small perturbations around $\varphi^*$.
	
	We can diagonalize the dynamics (\ref{eq:infinite_dim_linearized_dynamics}) by writing $\eta$ as a Fourier series,
	\begin{equation}
	\label{eq:eta_fourier}
	\eta(\omega) = \sum_{k\in\mathbb{Z}} c_k(t) e^{ik\pi\omega} \,.
	\end{equation}
	Inserting the form (\ref{eq:eta_fourier}) into the $\mathcal{O}(\epsilon^1)$ equation (\ref{eq:infinite_dim_linearized_dynamics}), we find that
	\begin{equation}
	\label{eq:fourier_mode_dynamics}
	\partial_t c_k = \left(\frac{-1}{\pi} + \delta_{|k|,1} \frac{K}{2} \right)c_k.
	\end{equation}
	
	All Fourier components of the perturbation $\eta$ except for the first decay exponentially in time with a rate $1/ \pi$, while the first Fourier component will grow or shrink with time, depending on the sign of $-1/\pi + K/2$. Specifically, if $K<2/\pi$, then the first Fourier mode also decays in time, while if $K>2/\pi$, the first Fourier mode grows in time, and the fixed point $\varphi^*$ is unstable. Hence we have, as in the finite-$N$ case, the critical coupling strength $K_c = 2/\pi$.
	
	\subsection{Interpretation} \label{subsec:interp}
	
	In both the finite- and infinite-dimensional versions of our model, we have found that nonzero synchrony spontaneously develops as the coupling strength $K$ exceeds $K_c = 2/\pi$. We contrast this result with that for the corresponding unforced model,
	\begin{equation}
	\dot{\varphi_i} = \omega_i + \frac{K}{N}\sum_{j=1}^{N}\sin(\varphi_j-\varphi_i) \,,
	\end{equation}
	where $\omega_i = 2\pi i / N -\pi$. While the standard result (\ref{eq:unforced_kc}) does not directly apply in this case, since the uniform density is not unimodal, it has been established by Paz\'{o} \cite{pazo2005} that the synchronization transition does in fact occur at $K_c^{\text{unf}} = 4/\pi = 2/(\pi g(0))$, which is twice the value at which the forced model begins to show nonzero synchrony. This result can be considered surprising, given that we have taken a forcing term, (\ref{eq:sawtooth_interaction_function}), that was designed specifically to drive the system to a state of zero synchrony. 
	
	The situation becomes clearer if we compare the desynchronized state present in the forced model to the incoherent state in the unforced model.  The desynchronized state, defined by $\varphi(\omega) = \pi \omega$, has zero synchrony as measured by the order parameter $R$. However, it has the property that every oscillator moves at equal frequency. This is in contrast with the incoherent state of the unforced Kuramoto model, in which each oscillator moves at its own natural frequency. Hence, in the sense of frequencies, the desynchronized state is far more organized than the incoherent state, although this fact is missed by the synchrony parameter $R$, which only measures instantaneous alignment of phases.
	
	To understand the role of frequency alignment in establishing phase alignment, it is instructive to consider again the standard unforced Kuramoto model. As we have already quoted (\ref{eq:unforced_kc}), the critical coupling strength is $K_c^{\text{unf}} = 2/\pi g(0)$, where $g$ is the density of the distribution of natural frequencies. Intuitively, this expression captures the trade-off between disorder in the natural frequencies and the ordering influence of coupling; the tighter the distribution of natural frequencies, the larger $g(0)$, and the smaller $K_c^{\text{unf}}$. In other words, the coupling strength must be large enough to overcome the diversity of natural frequencies in order to bring about a preferred phase.
	
	In the desynchronized state of the forced model, the oscillators move with a single frequency. Hence, there is no disorder to be overcome by the coupling. All that keeps the system in the desynchronized state is the forcing, which appears as the eigenvalue $-1/\pi$ in the spectrum of the Jacobian. The second eigenvalue, $-1/\pi + K/2$, directly captures the trade-off between the driving and the coupling, showing that the stability of the entrained state is the only force that needs to be countered by coupling.

	\subsection{On the Relation to Previous Work}
	\label{sec:relation_to_previous_work}
	
	Finally, we discuss the relationship of the present model to previous work on models of globally coupled oscillators subject to common forcing. The existing literature has focused almost exclusively on sinusoidal forcing \cite{sakaguchi1988,mirollo1990a,ott2008low,childs2008stability,antonsen2008external}, owing to the analytical progress that this assumption allows.
	
	One such model was discussed by Ott and Antonsen \cite{ott2008low} as a possible application of the powerful dimension reduction known as the Ott-Antonsen (OA) ansatz. While it is the case that the OA ansatz can describe the fixed point that we consider, the dynamics away from the fixed point do not leave the OA manifold invariant, precisely because the sawtooth forcing function we consider is not sinusoidal.
	
	Another system much more closely similar to ours is the ``random pinning'' model studied by Mirollo and Strogatz \cite{mirollo1990a}. The random pinning model consists of a system of $N$ spins, with each one pinned by an anonymous driving force to a particular (randomly chosen) phase. In explicit terms, the dynamics are
	\begin{equation}
	\label{eq:jump_bif_dynamics_finite_N}
	\dot{\varphi_i} = \sin(\alpha_i - \varphi_i) + \frac{K}{N}\sum_{j=1}^{N} \sin(\varphi_j - \varphi_i)\,,
	\end{equation}
	where $\{ \alpha_i \}$ are random quantities sampled from the uniform distribution on the unit circle. The only difference between this equation and the one that we study is the term $\omega_i - \varphi_i/\pi$ is replaced by $\sin (\alpha_i - \varphi_i)$. It remains the case that in the absence of coupling, each oscillator evolves according to an autonomous ODE on the unit circle with one stable fixed point, and that the state in which each oscillator is at its individual fixed point has $R\approx 0$.
	
	The authors proceed to present a continuum formulation of the dynamics (\ref{eq:jump_bif_dynamics_finite_N}) that is of the same form as (\ref{eq:infinite_dim_dynamics}); where we represent phase as a function of natural frequency, they represent phase as a function of pinning phase $\alpha$. Owing to the regularity of the sine function, it is possible to obtain precise analytical results on the existence, number, and stability of fixed points. Our formulation is not amenable to the same analysis, for the reason that the sawtooth forcing function we consider has infinitely many Fourier modes.
	
	\section{Numerical Simulations} \label{sec:numeric}
	Here we present some numerical studies of the dynamical system (\ref{eq:dynamics}), which confirm the bifurcation at $K=K_c=2/\pi$ and illuminate the system's behavior away from the bifurcation point. To serve as reference, we also present data from numerical solution of the system in the absence of forcing which is the Kuramoto model with evenly spaced natural frequencies.
	
\begin{figure}[htb]
		\includegraphics[width=\columnwidth]{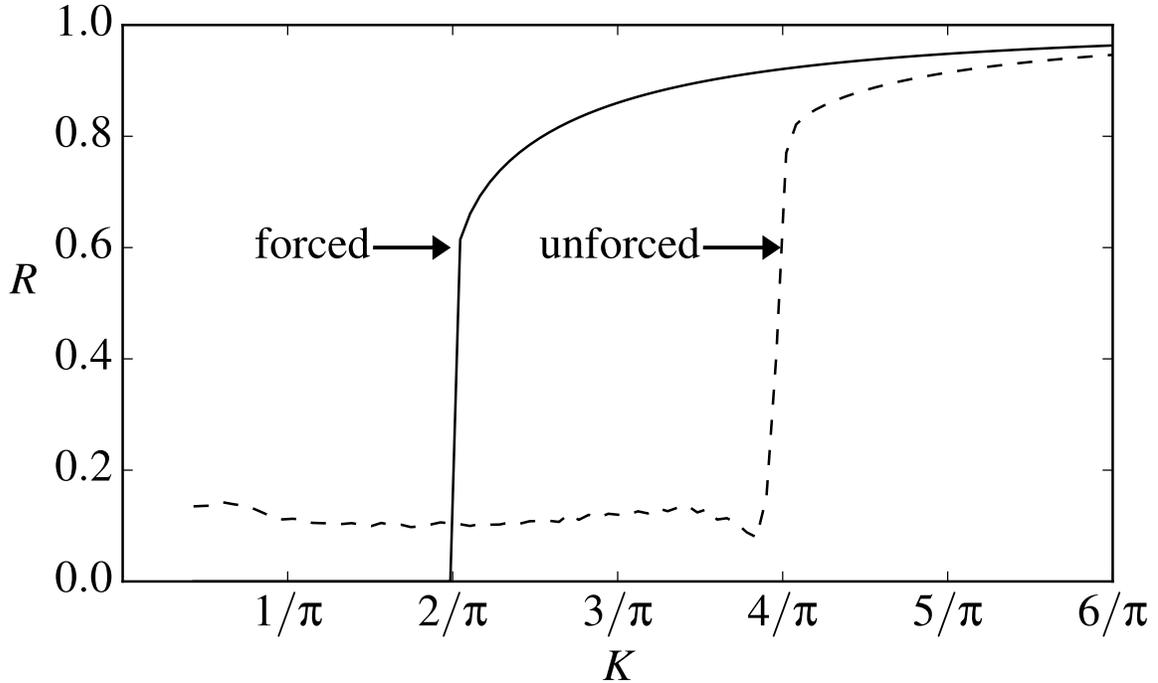}
		\caption{Synchronization $R$ vs.\ coupling strength $K$.
			In the unforced case (dashed line) the synchronization threshold is $K_c^{\text{unf}}=4/\pi$.  When forcing is added to drive the system to a splay state of equally distributed phase angles it synchronizes at a lower coupling strength $K_c=2/\pi$. The data were generated from a simulation of $N=50$ oscillators, starting from random initial conditions. For the forced case, integration was carried out until the system was determined to be at a fixed point. For the unforced case, integration was carried out until the system was determined to be in a statistically steady state.}
		\label{fig:general_sweep}
	\end{figure}

	As we can see in Fig.~\ref{fig:general_sweep}, the synchrony $R$ achieved at any value of the coupling strength $K$ greater than $2/\pi$ is greater in the forced case than in the unforced case. This confirms the conclusion that entrainment by broadcast periodic forcing has brought the system closer to synchrony, as measured by the order parameter $R$.
	
	In the numerical simulations above, we find a sharp increase in the steady-state value of $R$ as a function of coupling strength $K$. To obtain a deeper understanding of the nature of this transition and of the $R>0$ fixed point of (\ref{eq:dynamics}), we perform, for a range of $N$ values, numerical continuation of the $R>0$ fixed point with respect to the bifurcation parameter $K$ using the numerical continuation software AUTO \cite{Doedel07auto-07p:continuation}.
	
	To perform numerical continuation with AUTO, it is first necessary to locate an attractor (in our case, a fixed point) on the branch of interest. For each $N$ from 3 to 100, this was accomplished by numerical integration of (\ref{eq:dynamics}) until stationarity with $K=0.7$. This value of $K$ was chosen as it is greater than $K_c =2/\pi$, and was observed to lead to an $R>0$ fixed point in all instances. The AUTO software was then instructed to locate a connected family of fixed points in the joint parameter-state space $\mathbb{R}\times[-\pi,\pi]^N\ni(K,\varphi)$, searching in the negative $K$ direction from the user-supplied fixed point. AUTO equation and constants files, including initial fixed point locations for $3\le N\le 100$, are available upon request.
	
\begin{figure}[htb]
		\includegraphics[width = \columnwidth]{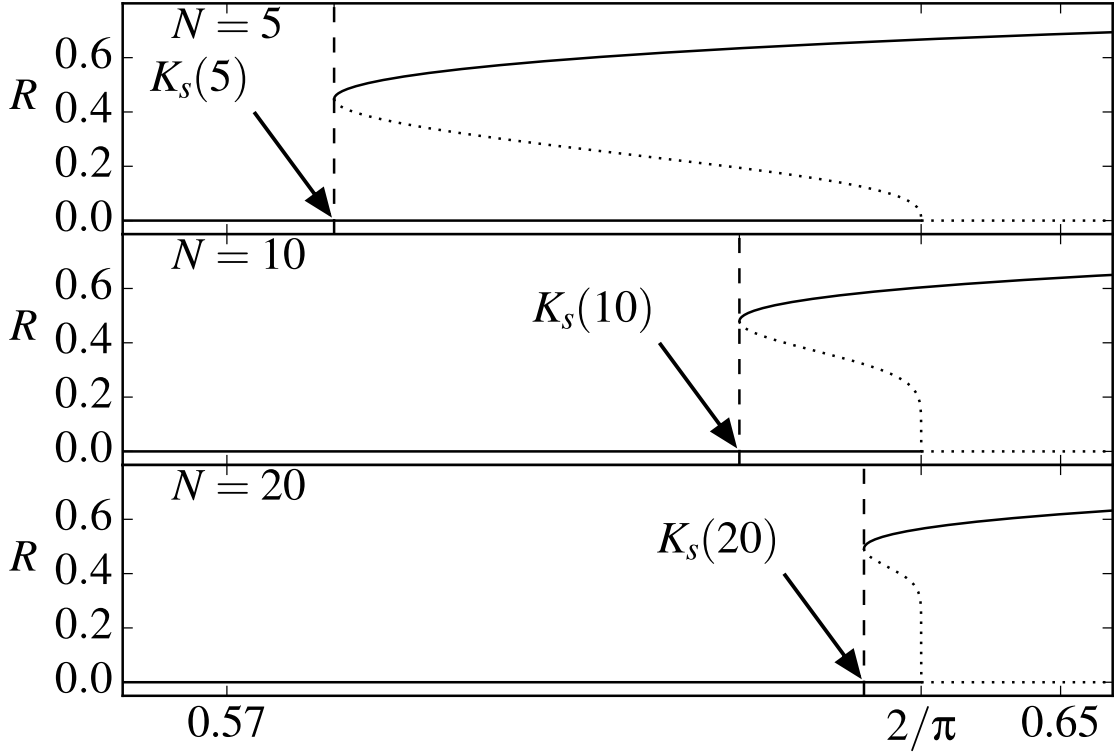}
		\caption{Bifurcation diagrams for the finite-$N$ system showing the bistable region as it depends on $N$. Solid lines indicate stable fixed points; dotted lines, unstable. Data generated using AUTO software\cite{Doedel07auto-07p:continuation}.}
		\label{fig:splitting}
	\end{figure}

	Representative results of the continuation just described are shown in Fig.~\ref{fig:splitting}. In particular, we find that for any $N=3\dots100$, the stable $R>0$ branch undergoes a saddle-node bifurcation at a coupling strength $K = K_s(N) < K_c = 2/\pi$. The unstable portion of this branch exists for all $K\in[K_s(N), K_c]$, and meets the $R=0$ branch (i.e. the desynchronized fixed point) transversally, precisely at $K=K_c$. Fig.~\ref{fig:splitting} clearly shows, for $N=5,10,20$, the existence of a bistable region $[K_s(N),K_c]$, implying that hysteresis is possible upon slow variation of $K$.
	
\begin{figure}[htb]
		\includegraphics[width=\columnwidth]{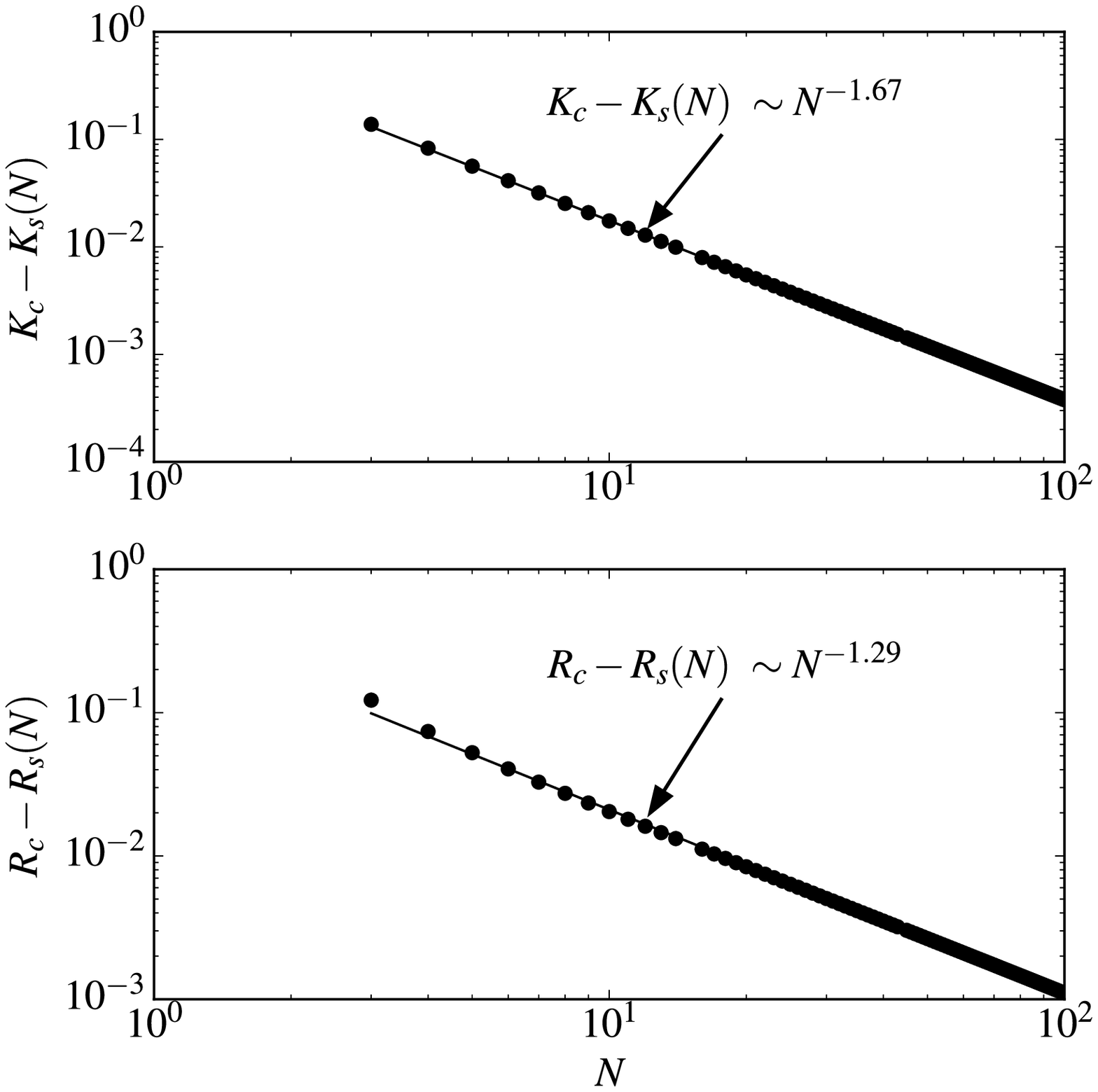}
		\caption{(upper) The extent of the stable $R>0$ branch below $K_c=2/\pi$ as a function of $N$. The data show approximately a power law scaling $N^{-1.67}$ for $N$ between $3$ and $100$.
			(lower) The difference between the value of synchrony $R$ at the saddle-node point and a numerically estimated critical value of $R_c = 1/2$ as a function of $N$. The data show approximately a power law scaling $N^{-1.29}$ for $N$ between $3$ and $100$.
			Circles represent data measured from AUTO simulation, solid line is a power law fit.
		}
		\label{fig:lower_lim_vs_N}
	\end{figure}

	Moreover, we find that the shape of the bifurcation diagram in the bistable region obeys a strong regularity across different values of $N$. In particular, the width of the bistable region, namely $K_c - K_s(N)$, follows a power-law scaling with $N$, with exponent $-1.67$. Additionally, the value of $R$ at the saddle-node point, which we denote $R_s(N)$, is observed to approach a value $R_c = 1/2$ from below, according to a power-law with exponent $-1.29$ (see Fig.~\ref{fig:lower_lim_vs_N}). We expect, therefore, that the infinite-$N$ system will exhibit a jump bifurcation at $K=2/\pi$ with a height (as measured by $R$) of $1/2$, but without a hysteresis loop.
	
	The situation is similar to that investigated by Paz\'{o} \cite{pazo2005}, who found the locking threshold for the (unforced) Kuramoto model with evenly spaced natural frequencies. In contrast with the typically considered case in which the density $g$ of the natural frequency distribution has $g''(0)<0$, leading to a continuous synchronization transition \cite{strogatz2000kuramoto}, the uniform distribution has $g''(0)=0$, and the transition is discontinuous. Precise results for the height of the jump, $R_c^{\text{unf}}$, and the scaling of $R-R_c^{\text{unf}}$ for $K>K_c^{\text{unf}}$ were derived using a self-consistent approach\cite{pazo2005}.
	
	Correspondingly, Paz\'{o} found, in the finite-$N$ system, a phenomenon of global frequency alignment for $K$ below the infinite-$N$ critical point, $K_c^{\text{unf}}$. Specifically, it happens that as coupling strength is increased, oscillators with nearby frequencies lock to each other, forming clumps, which then merge as $K$ is further increased. The final merge occurs at $K=K_s(N)$, which approaches $K_c^{\text{unf}}$ from below as $N\to\infty$, according to $K_c^{\text{unf}} - K_s(N) \sim N^{-\mu}$ with $\mu\approx 3/2$. We should note that for finite $N$, the transition in the unforced case is not hysteretic, as it is in the forced case.
	
	\section{Conclusions}
	
	We have explored, using an idealized model, the interplay between two ways in which a population of phase oscillators may be caused to behave coherently: common periodic forcing and attractive coupling. Based on the synchrony order parameter, $R$, forcing and coupling can appear to be at odds; the forcing drives $R$ towards zero while the coupling drives $R$ towards one. However, as we demonstrate both analytically and numerically, this view is inherently limited, since for $K$ above $K_c = 2/\pi$, the forced system exhibits greater phase alignment than the corresponding unforced system. An intuitive explanation for this mismatch is that the parameter $R$ measures only phase alignment, and is prone to miss the necessary precondition of frequency alignment.
	
	Though we have gained considerable intuition from the results already obtained, there is more work to be done. First, we are still lacking analytical understanding of the upper branch of solutions, which would include an expression for the height of the jump and the scaling of $R$ with $K$ above the jump (see Fig.~\ref{fig:splitting}).
	
	Another set of questions involves the (in)feasibility of the sawtooth interaction function $\Lambda_v$ (defined as $\Lambda_v(\varphi) = -\varphi / \pi$ for $\varphi \in (-\pi, \pi]$ (\ref{eq:sawtooth_interaction_function})). A simple argument reveals that for any integrable forcing waveform $v\in L^2(0,2\pi)$, the corresponding interaction function $\Lambda_v$ will be continuous on $S^1$, a condition which the sawtooth does not satisfy.  It remains unexplored to what degree the results presented here may be approximated by interaction functions that approximate a sawtooth. One could investigate the scaling of  dynamical properties with the energy of the input signal used.
	
	Techniques for analysis and control of entrainment processes can be used to examine and even manipulate numerous processes in biology \cite{tanaka2015entrainment}. In addition to developing an initial mathematical framework for characterizing stability of coherence phase structures in a continuum of interacting oscillators, our work presents a potential path towards addressing a compelling biological application.  Specifically, although some disagreement about the nature and phenomenology of epilepsy exists in the neuroscience literature \cite{jiruska2013synchronization}, studies in animal models have indicated that control of synchronization of neural dynamics can mitigate epileptiform activity \cite{good2009control}.  It is understood that neural stimulation is an underactuated system because one or a few electrodes are used to control the mean field of a very large collection of interacting neurons, which for practical purposes may be approximated by a continuum \cite{ching2012distributed}.  The ability to characterize the stability of phase decoherence in continuum models of general coupled oscillators could determine the possibility of developing effective desynchronizing stimuli for treatment of epilepsy.  The criterion that is derived in Section \ref{subsec:interp} and validated by numerical experiments in Section \ref{sec:numeric} could in principle be tested experimentally \cite{hunter03}.
	
	\begin{acknowledgments}
		
		This work was supported by the Laboratory Directed Research and Development program through the Center for Nonlinear Studies at Los Alamos National Laboratory under Department of Energy Contract No.\ DE-AC52-06NA25396. We gratefully acknowledge support from  the US Army Research Office MURI award W911NF-13-1-0340 and Cooperative Agreement W911NF-09-2-0053. The authors thank the anonymous reviewers for the constructive comments, and J.S. acknowledges Raissa D'Souza for useful discussions. 
		
	\end{acknowledgments}
	
	\appendix
	\section{Stability Analysis Calculations}
	\setcounter{section}{1}
	\subsection{Spectrum of the Jacobian in Finite Dimensions}
	\label{sec:fin_dim_jac}
	Here we calculate the spectrum of the $N\times N$ matrix $\mathbf{J}$ whose entries are
	\begin{equation}
	J_{ij} =\left(\Lambda'_{v}(\varphi_{i}^{*})-\frac{K}{N}\sum_{k\ne i}\cos(\varphi_{i}^{*}-\varphi_{k}^{*})\right)\delta_{ij}+ (1-\delta_{ij})\frac{K}{N}\cos\left(\varphi_{i}^{*}-\varphi_{j}^{*}\right)\,,
	\end{equation}
	where $\delta_{ij}$ is the Kronecker delta. By symmetry of the phase configuration, the sum in the diagonal term is independent of $i$, and can be computed by noticing
	\begin{align}
	0&=\sum_{k=1}^{N}\cos(\varphi_{i}^{*}-\varphi_{k}^{*}) \nonumber \\ &=\cos(\varphi_{i}^{*}-\varphi_{i}^{*})+\sum_{k\ne i}\cos(\varphi_{i}^{*}-\varphi_{k}^{*}) \,,
	\label{eq:symmetry_in_diag_of_J}
	\end{align}
	where the first equality in (\ref{eq:symmetry_in_diag_of_J}) follows from symmetry ($R=0$). Hence
	\begin{equation}
	\sum_{k\ne i}\cos(\varphi_{i}^{*}-\varphi_{k}^{*})=-\cos(\varphi_{i}^{*}-\varphi_{i}^{*})=-1 \,.
	\end{equation}
	This allows us to write the matrix entries in a simpler form, which will facilitate calculation of eigenvalues,
	\begin{align}
	J_{ij} &=\left(\Lambda'_{v}(\varphi_{i}^{*})+\frac{K}{N}\right)\delta_{ij}+(1-\delta_{ij})\frac{K}{N}\cos\left(\varphi_{i}^{*}-\varphi_{j}^{*}\right) \nonumber \\
	&= \frac{-1}{\pi}\delta_{ij} + \frac{K}{N}\cos(\varphi_i^*-\varphi_j^*) \,.
	\end{align}
	
	Note that we have used $\Lambda'_{v}(\varphi_i^*)=-1/\pi$ for all $i$, and $\delta_{ij} +(1-\delta_{ij}) \cos(\varphi_i^*-\varphi_j^*) = \cos(\varphi_i^*-\varphi_j^*)$ for all $i,j$. We can write this in matrix form as
	\begin{equation}
	\label{eq:jac_forced_matrix_form}
	\mathbf{J}=K\mathbf{C}-\frac{1}{\pi}\mathbf{I} \,,
	\end{equation}
	where $\mathbf{I}$ is the identity matrix and $\mathbf{C}$ is the matrix with entries $C_{ij}=N^{-1}\cos(\varphi_{i}^{*}-\varphi_{j}^{*})$. This form makes it clear (see calculation starting at (\ref{eq:spec_J_from_spec_C})) that to find the eigenvalues of $\mathbf{J}$ for arbitrary values of $K$, it suffices to find the eigenvalues of $\mathbf{C}$. To do this, we can write the action of $\mathbf{C}$ on an arbitrary vector $x$ as
	\begin{align}
	(\mathbf{C}x)_{i} = &  \sum_{j=1}^{N}C_{ij}x_{j}=N^{-1}\sum_{j=1}^{N}\cos(\varphi_{i}^{*}-\varphi_{j}^{*})x_{j} \nonumber\\
	= & \cos(\varphi_{i}^{*})\left[N^{-1}\sum_{j}\cos(\varphi_{j}^{*})x_{j}\right] \nonumber+\sin(\varphi_{i}^{*})\left[N^{-1}\sum_{j}\sin(\varphi_{j}^{*})x_{j}\right] \,,
	\end{align}
	where we have used the sum angle identity for cosine. The range of $\mathbf{C}$ is spanned by the vectors $e^{1}=(\cos(\varphi_{i}^{*}))_{i=1}^{N}$ and $e^{2}=(\sin(\varphi_{i}^{*}))_{i=1}^{N}$. Each of these is in fact an eigenvector with eigenvalue $1/2$, which follows from
	\begin{align}
	(\mathbf{C}e^{1})_i = & \cos(\varphi_{i}^{*})\left[N^{-1}\sum_{j}\cos(\varphi_{j}^{*})\cos(\varphi_{j}^{*})\right] \nonumber+\sin(\varphi_{i}^{*})\left[N^{-1}\sum_{j}\sin(\varphi_{j}^{*})\cos(\varphi_{j}^{*})\right] \nonumber\\
	= & \cos(\varphi_{i}^{*})\left[N^{-1}\sum_{j}\cos^2(\varphi_{j}^{*})\right] \nonumber =  \cos(\varphi_{i}^{*})\left[N^{-1}\sum_{j}\frac{1 + \cos(2\varphi_{j}^{*})}{2}\right] \nonumber\\
	= & \frac{1}{2}\cos(\varphi_{i}^{*}) = \frac{1}{2}(e^1)_i \,,
	\end{align}
	and similarly for $e^2$. Hence $e^{1}$ and $e^{2}$ are eigenvectors of $\mathbf{C}$ with eigenvalue $1/2$, and all other eigenvalues of $\mathbf{C}$ are zero.
	
	Finally, we can find the eigenvalues of $\mathbf{J}$ for arbitrary $K$. Notice that
	\begin{align}
	\label{eq:spec_J_from_spec_C}
	\lambda\in\sigma(\mathbf{J}) & \iff  \det(\mathbf{J}-\lambda\mathbf{I})=0 \nonumber\\
	& \iff  \det\left(K\mathbf{C}-\left(\frac{1}{\pi}+\lambda\right)\mathbf{I}\right)=0 \nonumber\\
	& \iff  \det\left(\mathbf{C}-K^{-1}\left(\frac{1}{\pi}+\lambda\right)\mathbf{I}\right)=0 \nonumber\\
	& \iff  K^{-1}\left(\frac{1}{\pi}+\lambda\right)\in\sigma(\mathbf{C}) \,.
	\end{align}
	Hence the eigenvalues $\lambda$ of $\mathbf{J}$ are of the form $\lambda=-1/\pi+K\mu$, for $\mu\in\sigma(\mathbf{C})=\left\{ 0,1/2\right\} $. In other words,
	\begin{equation}
	\sigma(\mathbf{J})=\left\{ \frac{-1}{\pi},\frac{-1}{\pi}+\frac{K}{2}\right\} \,.
	\end{equation}
	
	\subsection{Linearization in Infinite Dimensions}
	\label{sec:infinite_dim_linearize}
	
	Here we present the details of linearizing the infinite-dimensional dynamics (\ref{eq:infinite_dim_dynamics}) at the desynchronized fixed point $\varphi^*(\omega) = \pi \omega$.
	
	First, inserting the form $\varphi(\omega)=\varphi^*(\omega)+\epsilon\eta(\omega)$ into equation (\ref{eq:infinite_dim_dynamics}) yields
	\begin{align}
	\partial_t(&\varphi^*+\epsilon\eta) =  \omega -\frac{\varphi^*(\omega)+\epsilon\eta(\omega)}{\pi} + K\intop_{-1}^{1}\frac{1}{2} \sin\left( \varphi^*(\omega')-\varphi^*(\omega) + \epsilon\left(\eta(\omega')-\eta(\omega)\right) \right) d\omega' \,. \nonumber
	\end{align}
	Next we expand the sine function in the integrand around the point $\varphi^*(\omega')-\varphi^*(\omega)$, and obtain
	\begin{align}
	\sin(\varphi^*(\omega')-\varphi^*(\omega) + \epsilon(\eta(\omega')-\eta(\omega)) ) = &\nonumber \sin( \varphi^*(\omega')-\varphi^*(\omega))\nonumber \\
	+ &\epsilon\cos( \varphi^*(\omega')-\varphi^*(\omega))(\eta(\omega')-\eta(\omega)) \nonumber +  \mathcal{O}(\epsilon^2). 
	\end{align}
	
	From here we can read off the terms of order $\epsilon^0$ from each side of the equation, and get
	\begin{equation}
	\partial_t\varphi^* = \omega - \frac{\varphi^*(\omega)}{\pi} + K\intop_{-1}^{1} \frac{1}{2} \sin\left(\varphi^*(\omega')-\varphi^*(\omega)\right) d\omega' \,,\nonumber 
	\end{equation}
	which clearly holds, as each side evaluates to zero for all $\omega\in[-1,1]$.
	
	Next, we gather terms of order $\epsilon^1$ and obtain (dropping the $\epsilon$ factor from all terms)
	\begin{equation}
	\partial_t \eta = -\frac{\eta(\omega)}{\pi} \nonumber + K\intop_{-1}^{1}\frac{1}{2}\cos\left(\varphi^*(\omega')-\varphi^*(\omega)\right)\left[\eta(\omega')-\eta(\omega)\right] d\omega'.
	\end{equation}
	We can in fact simplify the integral above by noticing that
	\begin{align}
	\intop_{-1}^{1}\frac{1}{2}\cos\left(\varphi^*(\omega')-\varphi^*(\omega)\right)\left[\eta(\omega')-\eta(\omega)\right] d\omega' = \nonumber &\intop_{-1}^{1}\frac{1}{2}\cos\left(\varphi^*(\omega')-\varphi^*(\omega)\right)\eta(\omega') d\omega' \nonumber\\
	& - \eta(\omega)\intop_{-1}^{1}\frac{1}{2}\cos\left(\varphi^*(\omega')-\varphi^*(\omega)\right) d\omega' \nonumber \\
	= & \intop_{-1}^{1}\frac{1}{2}\cos\left(\varphi^*(\omega')-\varphi^*(\omega)\right)\eta(\omega') d\omega',
	\end{align}
	which follows from the symmetry of the phase configuration $\varphi^*$. We then arrive at the linearized dynamics as presented in the main text, (\ref{eq:infinite_dim_linearized_dynamics}), which we repeat here for completeness,
	\begin{equation}
	\label{eq:infinite_dim_linearized_app}
	\partial_t\eta(\omega) = -\frac{1}{\pi}\eta(\omega) + K\intop_{-1}^{1} \frac{1}{2} \cos(\varphi^*(\omega')-\varphi^*(\omega))\eta(\omega') d\omega'.
	\end{equation}
	
	Finally, we demonstrate the diagonalization of the linearized dynamics (\ref{eq:infinite_dim_linearized_app}) in the Fourier basis. As $\eta$ is a function on $[-1,1]$, the appropriate Fourier basis is $\{e^{ik\pi\omega}|k\in\mathbb{Z}\}$, so we write
	\begin{equation}
	\label{eq:eta_fourier_app}
	\eta(\omega) = \sum_{k\in\mathbb{Z}} c_k(t)e^{ik\pi\omega} \,,
	\end{equation}
	with the understanding that $\eta$ is real-valued and the coefficients $\{c_k\}$ will obey $\overline{c_k} = c_{-k}$, where the bar denotes complex conjugate.
	
	Next, we use $\varphi^*(\omega)=\pi \omega$ and Euler's formula to write
	\begin{equation}
	\label{eq:cosine_identity}
	\cos\left(\varphi^*(\omega')-\varphi^*(\omega)\right) = \frac{1}{2}\left( e^{i\pi(\omega'-\omega)} + e^{-i\pi(\omega'-\omega)}\right) \,.
	\end{equation}
	Inserting (\ref{eq:cosine_identity}) and (\ref{eq:eta_fourier_app}) into (\ref{eq:infinite_dim_linearized_app}) gives
	\begin{equation}
	\label{eq:infinite_dim_diagonal_app}
	\partial_t \eta(\omega) = -\frac{1}{\pi}\eta(\omega) + K\intop_{-1}^{1}\frac{1}{4}\sum_{k\in\mathbb{Z}} c_k(t)e^{ik\pi\omega'}\left(e^{i\pi(\omega'-\omega)} + e^{-i\pi(\omega'-\omega)}\right) d\omega'.
	\end{equation}
	The only terms of the sum that do not vanish in the integral are those with $k=\pm1$. For the $k=\pm1$ terms, the integral evaluates to
	\begin{equation}
	\intop_{-1}^{1}\frac{1}{4}c_1(t)e^{i\pi\omega'}e^{-i\pi(\omega'-\omega)}d\omega' = \frac{1}{2}c_1(t)e^{i\pi\omega}
	\end{equation}
	and likewise for $k=-1$. This shows that the coupling term acts on $\eta$ diagonally in the Fourier basis. Equating Fourier coefficients on each side of (\ref{eq:infinite_dim_diagonal_app}), we obtain
	\begin{align}
	k=\pm 1 \colon \quad \partial_t c_k(t) & = \left(\frac{-1}{\pi} + \frac{K}{2}\right) c_k(t) \\
	k\ne \pm 1\colon \quad \partial_t c_k(t) & = \frac{-1}{\pi} c_k(t) \,.
	\end{align}

	\section*{References}
	%
	
\end{document}